\documentclass[traditabstract,letter]{aa}

\usepackage{natbib}
\usepackage{epsfig}
\usepackage{epsf}
\usepackage{color}
\definecolor{red}{rgb}{0.7,0,0}
\definecolor{blue}{rgb}{0,0,0.7}
\def\correc#1{#1}

\usepackage{amssymb}
\usepackage{gensymb}

\def\ergcms{erg/cm$^{2}$/s}

\def\cm2{cm$^{-2}$}

\def\integral{{\it{INTEGRAL}}}

\def\chisq{$\chi^2_\nu$}

\def\v{V404}
\def\grs{GRS~1915+105}

\usepackage{longtable}
 \usepackage[figuresright]{rotating}
\usepackage{lscape}
\usepackage{txfonts}
\begin{document}
   \title{Correlated optical, X-ray, and $\gamma$-ray  flaring activity seen with \textsl{INTEGRAL} during the 2015 outburst of \v\ Cygni }
\subtitle{}

  \author{J. Rodriguez\inst{1},   M. Cadolle Bel\inst{2},   
  J. Alfonso-Garz\'on\inst{3},  T. Siegert\inst{4}, X.-L. Zhang\inst{4},  V. Grinberg\inst{5}, V. Savchenko\inst{6},  J. A. Tomsick\inst{7}, J. Chenevez\inst{8}, M. Clavel\inst{1}, S. Corbel\inst{1}, R. Diehl\inst{4}, A. Domingo\inst{3},   C. Gouiff\`es\inst{1}, J. Greiner\inst{4},  M. G. H. Krause\inst{9,4},  P. Laurent\inst{6}, A. Loh\inst{1}, S. Markoff\inst{10}, J. M. Mas-Hesse\inst{3}, J. C. A.  Miller-Jones\inst{11}, D. M. Russell\inst{12}, \and J. Wilms\inst{13}
   }

   \offprints{J. Rodriguez:  jrodriguez@cea.fr}
\authorrunning{Rodriguez  et al. }

\titlerunning{Correlated X/$\gamma$-ray/Optical flaring in \v}
 
  \institute{Laboratoire AIM, UMR 7158, CEA/CNRS/Universit\'e Paris Diderot, CEA DSM/IRFU/SAp, F-91191 Gif-sur-Yvette, France 
  \and  Max Planck Computing and Data Facility, D-85748 Garching, Germany  
  \and Centro de Astrobiolog\'{i}a (CSIC-INTA), ESAC Campus, POB 78, E-28691 Villanueva de la Ca\~nada, Spain 
  \and Max Planck Institut f\"ur extraterrestrische Physik, D-85748 Garching, Germany 
  \and Massachusetts Institute of Technology, Kavli Institute for Astrophysics and Space Research, Cambridge, MA 02139, USA 
  \and Laboratoire APC, UMR 7164, CEA/CNRS/Universit\'e Paris Diderot, Paris, France 
  \and Space Science Lab, University of California, Berkeley, CA 94720, USA 
  \and DTU Space -- National Space Institute, Technical University of Denmark, Elektrovej 327-328, 2800 Lyngby,  Denmark
  \and Universit\"ats-Sternwarte Muenchen, Ludwig-Maximilians-Universit\"at, Scheinerstr. 1, D-81679 Muenchen, Germany
  \and Anton Pannekoek Institute for Astronomy, University of Amsterdam, P.O. Box 94249, 1090 GE Amsterdam, the Netherlands
  \and International Centre for Radio Astronomy Research, Curtin University, GPO Box U1987, Perth, WA 6845, Australia
  \and New York University Abu Dhabi, PO Box 129188, Abu Dhabi, United Arab Emirates
  \and Dr. K. Remeis-Sternwarte \& Erlangen Centre for Astroparticle Physics,  Universit\"at Erlangen-N\"urnberg, Sternwartstr. 7, D-96049 Bamberg, Germany
}

   \date{}

 
  \abstract{After 25 years of quiescence, the microquasar V404 Cyg  entered a new period of activity in June 2015. This 
  X-ray source is known to undergo extremely bright and variable outbursts seen at all wavelengths. It is therefore 
  an  object of prime interest to understand the accretion-ejection connections. These can, however, only be  probed through simultaneous observations 
  at several wavelengths. We made use of the \integral\ instruments to obtain long, almost uninterrupted observations 
  \correc{from 2015 June 20$^{\mathrm{th}}$, 15:50 UTC to June 25$^{\mathrm{th}}$, 4:05 UTC,}  from the optical 
  V-band, up to the soft $\gamma$-rays. V404 Cyg was extremely variable in all bands, with the detection of 18 flares with fluxes exceeding 
  6 Crab (20--40 keV) within 3 days.  The flare recurrence can be as short as \correc{$\sim$}20~min from peak to peak. A model-independent analysis  shows that 
  the $>$6 Crab flares have a hard spectrum. A simple 10--400 keV spectral analysis of the off-flare and flare periods shows that the variation
  in intensity is likely to be due to variations of a cut-off power law component only. The optical flares 
 seem to be at least of two different types: one occurring in simultaneity with 
  the X-ray flares, the other showing a delay greater than 10 min. The former could be associated with X-ray reprocessing  by either
  an accretion disk or the companion star. We suggest that the latter are associated with plasma ejections that have also 
  been seen in radio. } 

   \maketitle
%

\section{Introduction}
\label{sec:intro}
\object{V404 Cyg}  (hereafter \v) is a low mass X-ray binary (LMXB) consisting of a black hole (BH)
with mass estimates ranging from $\sim$$9$ to $15$\,M$_\odot$, and a $0.7_{-0.2}^{+0.3}$\,M$_\odot$
K3 III companion \citep{Casares94,Shahbaz_94,Khargharia10}, located at
a parallax distance 2.39$\pm$0.14\,kpc \citep{MillerJ09}. The inclination of the binary's
rotational axis is $67\degree$$_{-1}^{+3}$ \citep{Shahbaz_94,Khargharia10}, the orbital period 6.5\,d
\citep{Casares92}. This transient underwent three periods of
outbursts during the 20th century \citep{richter89},  the last, in May 1989, leading
to its discovery as an X-ray transient by the Ginga satellite
\citep[as \object{GS 2023+338},][]{Makino89}.  \v\ showed bright X-ray
flares  on short time scales \citep[e.g.][]{Makino89,Terada94}, and this makes it an excellent source to study 
the connections between the accretion and ejection phenomena, which are the probable  origin of this behavior. 
\v\ is one of the closest stellar-mass BHs, making 
it a rare case where quiescence could be studied in detail, and variable remnant activity, 
attributed to a compact jet, was detected from radio to hard X-rays  \citep[e.g.][]{hynes04,xie14}. 
\v\ is one of the few sources that defines the radio/X-ray correlation over a large range of luminosities, down into quiescence 
\citep{corbel08b}. The good knowledge of the quiescent state makes understanding new outburst observations 
 paramount as they allow the mechanisms responsible for the increased activity to be probed.\\
\indent On 2015 June 15 (MJD 57188), \v\ went into outburst again. It was first
detected by {\it{Swift}} \correc{(BAT and XRT)} \citep{Barthelmy15}, and then with {\it{MAXI}}, and {\it{INTEGRAL}}
\citep{Negoro_Atel7646,Kuulkers_Atel7647}. These early alerts
triggered follow-up observations at all wavelengths.  Preliminary
results all report the detection of the source, variations of specific
spectral features, and an extreme flaring activity
at all wavelengths \citep[e.g.][]{Mooley_ATel7658,Munoz_ATel7659,Ferrigno_ATel7662,Motta_ATel7666,Tetarenko_ATel7661,Tetarenko_ATel7708,Tsubono_ATel7701}.  
 We triggered our \integral\ ToO program to obtain quasi-continuous
coverage in X/$\gamma$-rays and in the optical V band. The first detection of multiple flares 
exceeding 30\,Crab in 20--40\,keV, and possible correlated flaring in the V-band were reported in 
\citet{rodrigue_ATel7702}, and \citet{Domingo_ATel7717}. \\
\indent Our observations caught the source during the most intense and variable phase of this new outburst.  
Here, we first give the details of the observations and data reduction (Sec.~\ref{sec:data}). We then
focus on the flaring behavior at high energies (Sec.~3) that we compare to the activity in the optical 
V-band (Sec.~4). A preliminary phenomenological spectral characterization of different intensity intervals 
is presented in Sec.~5.   We discuss our results and compare \v\ to other microquasars in Sec. 6.
\begin{figure*}[!t]
\hspace*{-0.5cm}\epsfig{file=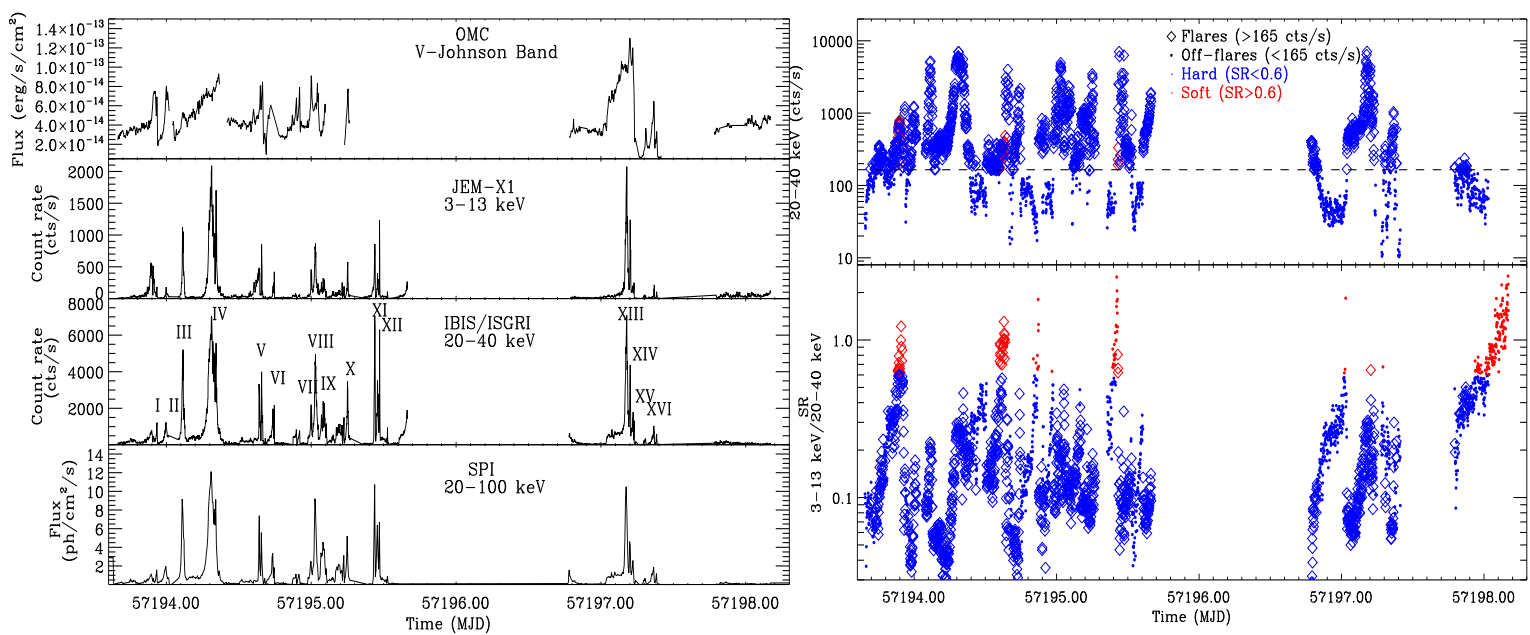,width=20cm}
\caption{ {\bf{Left:}} \integral\ LCs of \v\ in four spectral domains (a \correc{larger version of the plot including}  all energy ranges is available as online material). {\bf{Right:}} 20--40 keV count rate (top),  and 3--13/20--40 keV softness ratio (SR, bottom).  The dashed horizontal line corresponds to 1 Crab (20--40 keV). 
\correc{MJD 57193 is 2015 June 20$^{\mathrm{th}}$}.}
\label{fig:X}
\end{figure*}

\section{Observations and Data Reduction}
\label{sec:data}
Our ToO program (Fig.~\ref{fig:X}) covered MJD~57193.66--57198.17 
\correc{(2015-06-20 15:50 UTC to 2015-06-25 4:05 UTC)},
i.e., \integral\ revolutions \#1555 (continuous coverage) and \#1556 (two periods).
The data of all the \integral\ instruments \citep[see][and references therein for all instrumental details]{winkler03} 
were reduced with the {\tt{Off line Scientific Analysis  (OSA)}} v10.1 software suite, with the latest
calibration files available at the time of writing. \\
\indent Images and 100~s binned light curves (LC)  from the Joint European X-ray Monitors (JEM-X) and 
the Imager on Board the \integral\ Satellite (IBIS)  were produced in two bands (3--13, and 13--30\, keV) 
for JEM-X unit 1, and  in four bands (20--40, 40--80, 80--150, and 150--300\,keV) 
for the IBIS Soft Gamma-Ray Imager (ISGRI).  \\
\indent The event data of the Spectrometer on \integral\ (SPI) have been fitted with models for the celestial sources  
and instrumental background following standard reduction processes. The 20--100 keV LC of \v\, as well as the other sources
 in the field were obtained in bins of 400~s. 
  Background models have been built based on the pre-flaring data of a representative empty sky region, adjusting 
 the normalization coefficient per hour \citep[see, e.g.,][for a more general description of the method]{Strong05}.\\
\indent As source intensity and hardness vary strongly on short time scales, \correc{we extracted luminosity/hardness dependent JEM-X, ISGRI, and SPI 
spectra over specific time intervals of clean data. The} spectra from the same time intervals were jointly fitted 
within {\tt{XSPEC v12.8.2}}.  Since the instruments' responses are possibly different for the high intensities observed, only  
phenomenological spectral fits are presented and the fit results should be viewed with some caution. \\
\indent The \integral/optical Monitoring Camera (OMC)  fluxes and magnitudes were derived from a photometric aperture of $3 \times 3$
pixels (1 pix. =17.504$\arcsec$), slightly circularized, i.e. removing one
quarter pixel from each corner (standard output from {\tt{OSA}}). The 
photometric aperture was centered on the source coordinates (default
centroid algorithm) and did not include any significant contribution from other
objects. We removed measurements with a 
severe problem flag, and, to restrict the noise, only measurements of 50 and 200~s duration were considered.

\section{Model independent description of the flaring}
\label{sec:temporal}
Multi-wavelength LCs of \v\, from the V band up to 
$\gamma$-rays are highly structured with several large flares separated by calmer periods seen in all bands (Fig.~\ref{fig:X}, and 
see also Fig.~\ref{fig:onlineMulti} for a plot with all energy ranges).
In the following,  count rates (CR) are given in the \correc{ISGRI} 20--40\,keV range. \correc{When the source 
CR increased above $\sim$150--200 cts/s an intense X-ray flare systematically followed. In the following, we thus set} 
1~Crab\footnote{The ISGRI/20--40 keV CR of the Crab is 165 cts/s  $\Leftrightarrow F_{20-40{\mathrm{keV}}}$=8~$10^{-9}$~\ergcms\ 
for a  power law spectrum with $\Gamma$=2.1 and a normalization of 10  ph/cm$^{-2}$/s at 1~keV. }  as the typical limit
between \correc{the off-flare} and flaring \correc{intervals.} 
We identify 18 main events, i.e. peaks that reached at least 6~Crab (labeled with Roman numerals\footnote{V and XII
contain two distinct events hardly distinguishable in Fig.~\ref{fig:X}. They appear under the same label in Fig. 1 (to keep it clear), and 
are named with a/b sub-labels in the text and table~\ref{tab:flares}.} in Fig.~\ref{fig:X}, their 
main characteristics are given in the online table \ref{tab:flares}), with 
11 exceeding 20~Crab during our observations. \correc{Flares IV, XI, and XIII} are the brightest we observed, 
reaching \correc{43}~Crab.  The flares occur isolated (e.g. III, IV, VI) 
as well as in groups with peak to peak intervals as short as \correc{22~min (Va, Vb)}
The flares last \correc{0.4--2.4~hr} except peaks IV and XIII. 
The former shows a rather broad profile and has multiple peaks. This
 event lasted 4.8~hr in total and it is the longest flare of our
observation. The latter reached about 40 Crab\correc{.The peak itself lasted about 1.5~hr, but was preceded by a $\sim$3~hr long,  
3~Crab plateau seen only above 13 keV}. It was followed 
by flares XIV and XV which show decreasing peak values.  \\
\indent The 3--13/20--40 keV softness ratio (SR, Fig.~\ref{fig:X}, right) shows that the large variability of the source is  
associated with variations of SR from $\sim$0.03 to $\sim$1.3 \correc{corresponding to $\Gamma$=0.1--2.5 in simulated JEM-X/ISGRI 
power law spectra. SR$\sim$0.6 corresponds to $\Gamma$=2.0.}
\correc{Large spectral variations are visible in the off-flare intervals} (Fig.~\ref{fig:X}, right).
All the flares are hard, \correc{and they} even all have SR$<$0.4 ($\Leftrightarrow$$\Gamma$$<$1.8).   

\section{Optical vs. X-ray behavior}
\begin{figure}[!t]
\centering
\epsfig{file=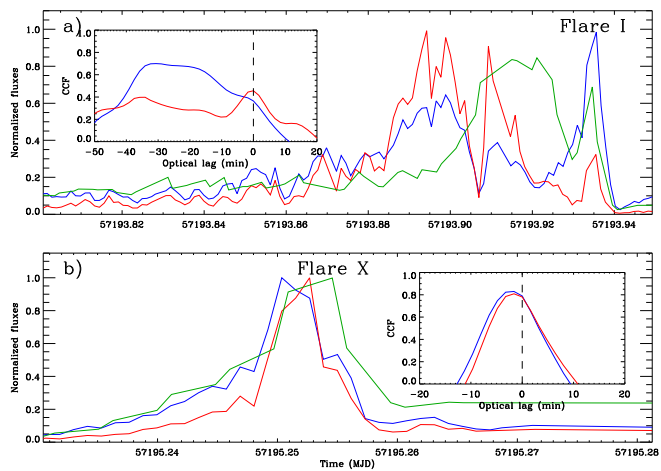, width=\columnwidth}
 \caption{\correc{V-band (green), 3--13 keV (red), and 20--40 keV (blue) LCs around flares I (a) and X (b)}. 
 The inserts show the cross-correlation functions of \correc{the 3--13 keV (red) and 20--40 keV (blue) vs the optical} LCs  
 \correc{over the same time intervals as the LCs}. 
 \correc{The dashed vertical lines represent the 0 lag level}.}
\label{fig:OptX}
\end{figure}

\indent The comparison of the optical (OMC) and X-ray (JEM-X1 and ISGRI) LCs shows a non trivial relationship.  
Significant flaring activity is  evident in the V band LC (Fig.~\ref{fig:X}, left), \correc{with at least 12 clear flares. The optical flare typically last 0.24--2.5~hr}. 
While some events \correc{occur in simultaneity with X-ray flares}, the optical emission is delayed with respect to the X-rays in other cases. 
\correc{Fig.~\ref{fig:OptX} shows typical examples of these different behaviors.} 
The cross-correlation function \correc{(ccf)} between the X-ray  and optical emission  confirms the absence of  lags for some of 
the flares \correc{(e.g. flare I that causes the peak at 0 in the ccf of Fig.~\ref{fig:OptX}a), and } delayed optical emission from 
1.5~min to \correc{20--}30~min is seen in others. \correc{The ccf of Fig.~\ref{fig:OptX}b shows an example of a $\sim$3~min lag, while the 
ccf of Fig.~\ref{fig:OptX}a (in addition to showing the simultaneity of peak I and its optical counterpart)  shows lag at 20--30~min 
representing the delay between  the small X-ray flares preceding peak I (around MJD 53193.9) and the subsequent optical flare (around MJD 53193.92).} 
Due to the time resolution of OMC, however, lags shorter than 1\, min can not be measured with our data, and additional lags of the order of seconds are not excluded.\\
\correc{\indent While most of the flares show a fast rise similar to the flares observed in X-rays, the two optical events occurring close to X-ray
  peaks IV and XIII seem to be exceptional. All other flares show fast rises ($\lesssim$1~hr), but 
these two events have slower rises (about 10 and 4~hr, respectively), and are both coincident with hard plateaus that precede the X-ray peaks. } 

\section{Spectral analysis}
\label{sec:spectral}
\begin{figure}[htbp]
\epsfig{file=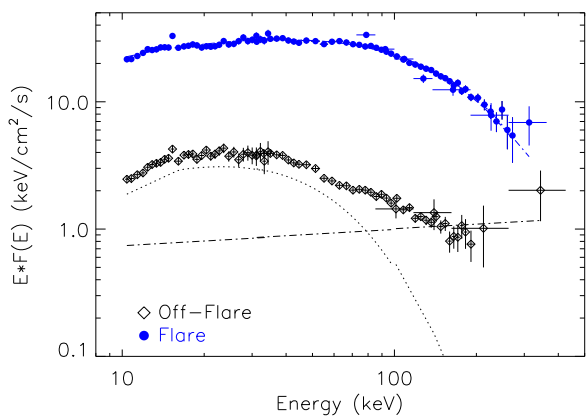,width=8.5cm}
\caption{$\nu$-f$_\nu$ spectra obtained from the fits to the \correc{off-flare} and flare intervals. The lines represent the 
spectral components used for the fits.}
\label{fig:Spec}
\end{figure}

\correc{We accumulated spectra from the brightest flares (CR$>$1000~cts/s) and the off-flare intervals (CR$<$165~cts/s). In the latter case we 
also only retained the hard intervals (SR$<$0.6) in order to exclude the softening visible after MJD 57197.9 (Fig.~\ref{fig:X}).}   
The resultant  ``$\nu-$f$_{\nu}$'' spectra are plotted in Fig.~\ref{fig:Spec}.\\
\indent The \correc{off-flare} spectrum is \correc{well fitted (\chisq=1.2, 66} degrees of freedom, dof) by a model consisting of a  
power law with a high-energy 
cut-off dominating at 10--100 keV plus  an additional power law dominating above  100~keV.
The former component has $\Gamma$=\correc{1.0$_{-0.4}^{+0.3}$, E$_{\mathrm{cut}}$=16$_{-2}^{+4}$~keV, E$_{\mathrm{fold}}$=23$\pm5$~keV}, 
the latter has \correc{$\Gamma$=1.9$_{-0.3}^{+0.2}$}, \correc{($\Gamma$ is the photon index defined as N(E)$\propto$E$^{-\Gamma}$}.) 
Normalization constants were included to account for potential cross-calibration issues, or differences in the effective  
exposures (deadtime corrections, telemetry drop out). When set 
to 1 for ISGRI, we get $\sim$1.9 for SPI and $\sim$0.6 for JEM-X1. 
The 20--400~keV \correc{(ISGRI)} flux is $\sim$10$^{-8}$\, \ergcms, and the above model leads to an extrapolated 0.1--10$^5$~keV flux
 $\sim$3.8 10$^{-8}$\, \ergcms, i.e. about 2\%\, L$_{Edd}$ for a 9~M$_\odot$ BH.  \\
 \indent The flare spectrum is well represented (\chisq$\sim$0.9, 78 dof) by a 
 \correc{single} cut-off  power law \correc{with} $\Gamma$=1.54$\pm$0.06,  E$_{\mathrm{cut}}$=14.0$_{-3.3}^{+2.8}$~keV, E$_{\mathrm{fold}}$=87$_{-5}^{+4}$~keV.  \correc{An extra power law component is not statistically required 
 according to an F-test. The normalization constants are both close to 1.1}  
  The 20--400~keV \correc{(ISGRI)} flux is $\sim$10$^{-7}$\ergcms, which  
 leads to an extrapolated 0.1--10$^5$~keV flux $\sim$3 10$^{-7}$\ergcms , i.e. about 20\% L$_{Edd}$ for a 9~M$_\odot$ BH.

\section{Discussion}
Over the 4 days covered by our \integral\ ToO, \v\ showed a high level of emission with sporadic flares 
 \correc{with a maximum 20--40~keV dynamical range of 940 (flare XVI)}. During its flares \v\ became the brightest X-ray object in the sky.  
In the hard \correc{off-flare} state the spectral analysis shows the presence of two spectral components: a cut-off power law 
usually attributed to thermal Comptonization and an extra power law at  energies beyond 100 keV. 
Hard tails have now been seen in a large number of systems \citep[e.g. GRS~1915+105, Swift J1753.5$-$0127, GX 339$-$4, or Cyg X-1,][]{rodrigue08_1915b,John15_tempo, Joinet07, rodrigue15_Cyg} and their origin is still highly debated, although 
a compact jet origin is favored in the case of \object{Cyg X-1} \citep[e.g.][]{russell14,rodrigue15_Cyg}. 
\correc{It is interesting to remark that the flaring activity seems primarily} due to spectral variations of the cut-off power law 
only.  We estimate an integrated luminosity L$\sim$\, 0.2L$_{\mathrm{Edd}}$ for the $>$6 Crab flares.  
 This is a lower limit to the maximum luminosity reached at the peak of the brightest flares. First because we averaged the data, without isolating 
 the brightest portions of the flares. Also we did not consider all the contributions  below 10 keV (disk, jet) that can provide a significant fraction of 
 the bolometric luminosity.  Assuming a simple scaling between the CR and L$_{\mathrm{Edd}}$ with a constant shape for the variable component
 between the \correc{off-flare} hard state and the flares,  we conclude that all peaks with CR$\gtrsim$3000 cts/s ($\sim$18 Crab)
(flares III, IV, Va,b, VIII, X, XI, XIIa,b, XIII and XIV, Fig.\ref{fig:X}) reached $\sim$L$_{\mathrm{Edd}}$. \\
 \indent The optical activity \correc{is also highly variable and shows flares}  (Fig.~\ref{fig:X}, \ref{fig:OptX}). 
Some optical flares occur in conjunction with the X-rays, while other activity periods show delays (Fig.~\ref{fig:OptX}). The first
specific length scale of this system is the separation between the BH and the companion: using the system parameters from Sec.\,\ref{sec:intro} and  
 a  9~M$_\odot$ BH we estimate  $\sim$2.2  $10^{12}$\,cm or $\sim$$75$\,light
seconds.  Hence, when no delay between the optical and the X-rays is observed (e.g flare I in Fig.~\ref{fig:OptX}a), the mechanism producing the
optical emission could be related to X-ray reprocessing, either by an accretion disk or by the companion. The maximum delays expected 
would be around 60~s (outer disk), and $\lesssim$150~s  (companion located at superior conjunction).  \\
\indent  Optical lags $\gtrsim$10\,min  could be related to variable jet properties, either as their intrinsic synchrotron emission, or from their 
interaction with the surrounding medium. Radio, and millimeter (mm) flaring activity ascribed to discrete ejections have been reported during 
this outburst \citep[e.g.][]{Sivakoff_ATel7671,Mooley_ATel7658, Tetarenko_ATel7708}. 
Delays between X-rays  and longer wavelengths are expected in the case of adiabatically expanding ejecta  \citep{vanderlaan66, Mirabel98}, 
and we remark that  \citet{Tetarenko_ATel7708} report a mm peak on MJD 57195.548, about 26 min after the 
MJD 57195.53 X-ray event  (a $<$6 Crab flare that occurred slightly after flare XII). 
Interestingly,  this delay also corresponds to the timescale of the mm flare increase \citep{Tetarenko_ATel7708} which renders  
the mm compatible with being causally related to the X-ray flare. As in  other well known sources, e.g.  \object{XTE J1550$-$564}, 
\object{GRO J1655$-$40}, \object{GRS 1915+105} \citep{Fender06b}, in \v\ discrete ejections may be causally connected to the X-ray activity.\\
\indent  \v\  reached maximum absolute peak intensities that are rather usual during
outbursts of  microquasars. \correc{If \v\ was at 10~kpc the maximum peak intensity would have been 2.5 Crab, a value} similar 
to the aforementioned other microquasars during their brightest outburst(s) \citep[e.g.][]{Remillard06}. 
Short recurrent (multi-wavelength) flares have, however, been 
seen only in GRS 1915+105  \citep[e.g.][]{Greiner96}, and in \v\ the flaring activity similarly \correc{recurs} on timescales as short as \correc{22~min}.  
Some of the \v\ optical flares lag \correc{$\gtrsim$20} minutes behind the X-rays (\correc{Fig.~\ref{fig:OptX}a}), and \correc{similar} lags are also seen 
at mm-radio wavelengths. \correc{This may resemble the} correlated X-ray/infrared/radio oscillations also 
referred to as  ``30-minutes" cycles of \grs\, \citep[e.g.][]{fender98,Mirabel98,rodrigue08_1915a}. In \grs, however, these events 
are associated with \correc{hard X-ray dips preceding the flares and} a clear softening at the X-ray peak, marking 
the disappearance of the Comptonization component \citep{rodrigue08_1915b}. 
 \v, on the contrary, remains hard even in the flaring states (Fig.~\ref{fig:X}), indicating a different mechanism responsible for the X-ray flaring 
 \citep[similar results were obtained from the 1989 outburst, e.g.][]{zycki99}.
 One tempting possibility would be that the high-energy flares are direct boosted emission from a jet (blazar-like configuration). This would imply
 a jet axis not perpendicular to the orbital plane. 
 Misaligned jets  have been seen  in GRO J1655$-$40 and \object{V4641 Sgr}  
\citep[][ and references therein]{maccarone02b}. In the former, a rather modest Lorentz factor $\gamma$$\sim$2.5 implies
small relativistic boosting \citep{hjellming95}. In the latter, $\gamma$ ranges from 10 up to 17, and the angle between the jet axis and the 
orbital plane normal is as high as 50$^\circ$ \citep[][ and references therein]{maccarone02b}. Significant Doppler boosting is expected in this case.\\
\indent \citet{zycki99} argued that the spectral and intensity variability seen with Ginga in 1989
could be due to the evolution of a heavily absorbing medium. However, even  with $N_H$ $\gtrsim$10$^{24}$\, cm$^{-2}$ \citep{zycki99},  
the activity above 20~keV is not affected by absorption, and hence the absorber alone cannot be responsible
for the large variability we observe. The high energy flares could be due to the shock of the relativistic 
jets with the dense ambient medium. Then optically thin synchrotron emission would be expected at X-ray energies, while our analysis favors
thermal Comptonization. 
 More simultaneous multi-wavelength observations will help disentangling these different possibilities.  
\begin{acknowledgements}
We warmly thank the referee for his/her useful report that helped us to improve the quality of this paper. 
We also thank the \integral\ teams and planners for their prompt reaction and the 
scheduling of these observations. JR,  MC, SC acknowledge funding support from the French Research National Agency: CHAOS project ANR-12-BS05-0009 (\texttt{http://www.chaos-project.fr}), and from the UnivEarthS Labex program of Sorbonne Paris Cit\'e (ANR-10-LABX-0023 and ANR-11-IDEX-0005-02). XLZ acknowledges funding through DLR 50 OG 1101. MGHK was supported by  the Deutsche Forschungsgemeinschaft under DFG project number PR 569/10-1 in the context of the Priority Program 1573 Physics of the Interstellar Medium. This work was supported by NASA through the Smithsonian Astrophysical
 Observatory (SAO) contract SV3-73016 for the Chandra X-Ray Center and
 Science Instruments. RD and XLZ acknowledge support through the Deutsches Zentrum 
  f\"ur Luft- und Raumfahrt e.V. (DLR) 50 OG 1101. OMC activities are supported by Spanish MINECO grant ESP2014-59789-P. This study is based on observations made with
 \integral, an ESA project with instruments and science data centre funded by ESA member states, Poland and with the participation 
 of Russia and the USA.  
\end{acknowledgements}


\Online
\onltab{
\begin{table*}
\caption{List of the $>$6~Crab flares and their main properties. \correc{MJD 57193 is 2015 June 20$^{\mathrm{th}}$}.}
\label{tab:flares}
\begin{tabular}{|l|ccc|cc|l|}
Name & Start$^a$ &  Peak time & Stop$^a$ & CR$_{3-13~{\mathrm{keV}}}$$^b$& CR$_{20-40~{\mathrm{keV}}}$$^b$ & Properties$^c$ \\
           & (MJD) & (MJD) & (MJD) & (cts/s) & (cts/s)  & \\
\hline
I & 57193.9217  & 57193.9356 & 57193.9402 &  184 & 1215 & Multiple \\
II & 57193.9703 &   57193.9981 & 57194.0827  & 181 & 1234 & Multiple \\
III & 57194.0827  & 57194.1152 & 57194.1428 & 1055 & 5209 & Complex \\
IV & 57194.2232 & 57194.3107  & 57194.3938 & 2010 & 7040 & Isolated/Complex\\
Va$^d$ &57194.6273 & 57194.6399 & 57194.6521 & 473$^e$ &3328 & Multiple/Complex, preceded by plateau \\
Vb$^d$ & 57194.6521 & 57194.6579 & 57194.6708 & 852 & 3999 & Multiple/Complex\\
VI & 57194.6960 & 57194.7346 & 57194.7473 & 129 & 1974 & Isolated/Complex \\
VII & 57194.9788 &  57194.9996 & 57195.0089 & 459 & 2200 & Multiple/Complex\\
VIII & 57195.0089 & 57195.0293 & 57195.0501 &  865 & 4950  & Multiple/Complex\\
IX & 57195.0582 & 57195.0826 & 57195.1095 & 320 & 2386 & Multiple/Complex  \\ 
X & 57195.2318 & 57195.2503& 57195.2712$^f$ & 577$^g$ & 3472 & Multiple, preceded by succession of $\sim$6 Crab peaks\\ 
XI & 57195.4294 & 57195.4388 & 57195.4450 & 857 & 7036 & Multiple \\
XIIa$^d$ & 57195.4450 & 57195.4573 & 57195.4665 & 401 &3525& Multiple/Complex\\
XIIb$^d$ & 57195.4665 & 57195.4723 & 57195.4841 & 1231 & 6299 & Multiple, followed by  plateau\\
XIII & 57197.1373 & 57197.1785 & 57197.1924 &2076 & 7081 & Multiple,  preceded by plateau\\
XIV & 57197.1924 & 57197.2020 & 57197.2067 &1240 & 4368 & Multiple \\
XV & 57197.2124 & 57197.2228 & 57197.2310 & 210 &1793 & Multiple/Complex\\
XVI & 57197.3450  &  57197.3647 & 57197.3705 & 151 & 1036 & Isolated\\
\hline
\end{tabular}
\begin{list}{}{}
\item[$^a$]Start (resp. stop) time of a flare is defined as the time 20--40 keV CR reaches 165~cts/s (1 Crab) during the increase (resp. decrease), or 
by the minimum reached before (resp. after) the increase (decrease) for multiple flares. The uncertainty on the times is $\pm$6$\times10^{-4}$~d.
\item[$^b$]Count rates at the peaks
\item[$^c$]``Multiple" stands for series of well defined flares occurring in rapid repetition. ``Complex" stands for flares showing various peaks. ``Plateau" indicates a $>$1 Crab plateau. 
\correc{\item[$^d$]These peaks appear as single peaks in Fig.~\ref{fig:X}. They are in fact true multiples.}
\correc{\item[$^e$]The 3--13 keV peak time occurred about 200\, s before the 20--40 keV one, indicating a potential hard lag.}
\correc{\item[$^f$]Data gap at the end of the flare. The stop time is the last point before the gap.}
\item[$^g$]The 3--13 keV peak time occurred about 200\, s after the 20--40 keV one, indicating a potential soft lag.

\end{list}
\end{table*} 
}

\onlfig{
\begin{figure*}
\centering
\epsfig{file=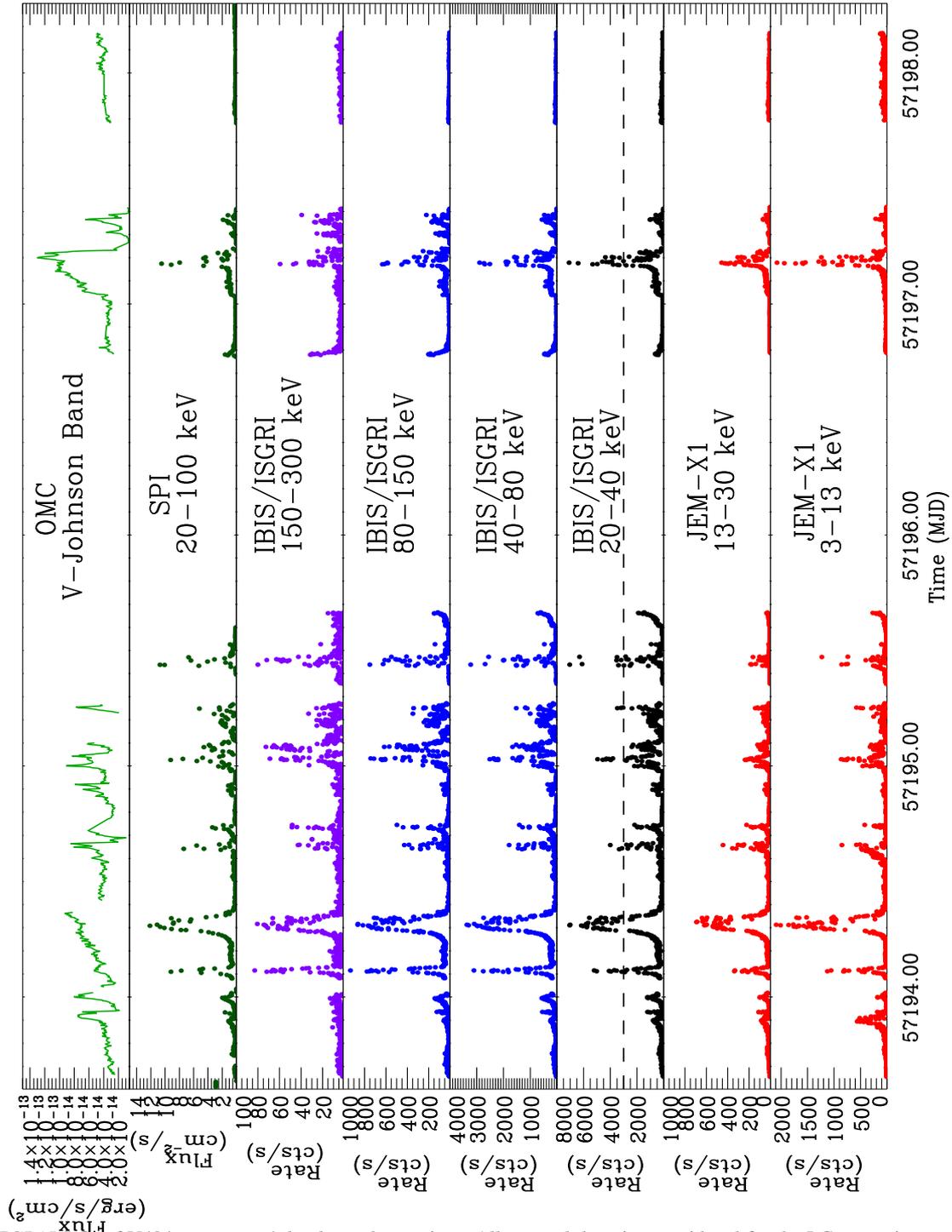, angle=90,width=15cm, height=19cm}
\caption{\integral\ LCs of \v\ over our $\sim$4 day-long observations. All spectral domains considered for the LC extraction are shown here. The dashed line 
in the 20--40 keV panel represents the approximate level of L$_{\mathrm{Edd}}$ we estimated. \correc{MJD 57193 is 2015 June 20$^{\mathrm{th}}$.}}
\label{fig:onlineMulti}
\end{figure*}
}
\end{document}